\documentclass[aps,prb,preprint,groupedaddress]{revtex4-1}
\usepackage{graphicx}
\usepackage{amsmath}
\usepackage{amsfonts}
\usepackage{textcomp}

\begin{document}
\newcommand{\guido}[1]{[\textbf{GUIDO:} \emph{#1}]}
\newcommand{\miquel}[1]{[\textbf{MIQUEL:} \emph{#1}]}
% Use the \preprint command to place your local institutional report
% number in the upper righthand corner of the title page in preprint mode.
% Multiple \preprint commands are allowed.
% Use the 'preprintnumbers' class option to override journal defaults
% to display numbers if necessary
%\preprint{}

%Title of paper
\title{Aharonov-Bohm oscillations and electron gas transitions in hexagonal core-shell nanowires with an axial magnetic field}

\author{Miquel Royo$^{1,2}$}
\email{mroyo@qfa.uji.es}
\author{Carlos Segarra$^{1}$}
\author{Andrea Bertoni$^{2}$}
\author{Guido Goldoni$^{2,3}$}
\author{Josep Planelles$^{1}$}

\affiliation{$^{1}$Departament de Qu\'{\i}mica F\'{\i}sica i Anal\'{\i}tica, Universitat Jaume I, E-12080, Castell\'o, Spain}
\affiliation{$^{2}$CNR-NANO S3, Institute for Nanoscience, Via Campi 213/a, 41125 Modena, Italy}
\affiliation{$^{3}$Department of Physics, Informatics and Mathematics, University of Modena and Reggio Emilia, Italy}

\date{\today}

\begin{abstract}
We use spin-density-functional theory within an envelope function approach to calculate  electronic states in a GaAs/InAs core-shell nanowire pierced by an axial magnetic field.
Our fully 3D quantum modeling includes explicitly the description of the realistic cross-section and composition of the sample, and  the electrostatic field induced by external gates in two different device geometries, gate-all-around and back-gate. 
At low magnetic fields, we investigate Aharonov-Bohm oscillations and signatures therein of the discrete symmetry of the electronic system, and we critically analyze recent magnetoconductance observations. 
At high magnetic fields we find that several charge and spin transitions occur. We discuss the origin of these transitions in terms of different localization and Coulomb regimes and predict their signatures in magnetoconductance experiments. 
\end{abstract}

% insert suggested PACS numbers in braces on next line
\pacs{}
% insert suggested keywords - APS authors don't need to do this
%\keywords{}

%\maketitle must follow title, authors, abstract, \pacs, and \keywords
\maketitle

% body of paper here - Use proper section commands
% References should be done using the \cite, \ref, and \label commands

\section{Introduction \label{intro}}

Gated semiconductor nanowire (NW) devices represent flexible test beds to study transport phenomena in the quasi-1D quantum regime. In this context, InAs-based NWs offer privileged properties derived, for instance, from the light InAs electron effective mass, which allows the experimental observation of the subband spectrum quantization even in NWs of relatively large section,\cite{ChuangNL13,VigneauPRL14,HalpernNL15} or from its large spin-orbit interaction and Land\'e factor,\cite{LiangNL12, BringerPRB11} which boost their prospectives in spintronics,\cite{NadjPergeNAT19} even at relatively high temperature.\cite{RossellaNNANO2014}
Furthermore, in this narrow gap material the Fermi energy, $E_F$, is pinned by surface states above the conduction band edge,\cite{LuthBook01} leading to an accumulation of electrons at the NW surface and facilitating the fabrication of ohmic contacts.\cite{ChuangNL13,BlomersNL11}

The resulting tubular shape of the conducting channel points toward interesting quantum phenomena under external magnetic fields.\cite{FerrariNL09} In particular, an axial field may lead to Aharonov-Bohm (AB) field-periodic modulation of the electron energy spectrum\cite{PlanellesPRB01} and, if the phase coherent length exceeds the perimeter of the NW, the observation of magnetoconductance oscillations.\cite{TserkovnyakPRB06,RosdahlPRB14} Indeed, several observations of AB-like oscillations in magnetotransport experiments performed on radial heterostructures have been reported.\cite{JungNL08,RichlerNL08,GulPRB14,GulNL14,WenzAPL14} Recently, G\"ul \emph{et al.}\cite{GulPRB14} have observed flux-periodic magnetoconductance oscillations in GaAs/InAs core-shell NWs. The oscillations persisted at different density regimes, modulated by a back-gate, exhibiting phase shifts as the back-gate voltage was gradually increased. A field-periodic magnetoconductance has also been observed in the same system with superconductor contacts\cite{GulNL14} and, after removal of the GaAs core, in a hollow InAs shell.\cite{WenzAPL14}

The single-crystal NW-based heterostructures investigated in these experiments have a prismatic hexagonal cross-section. However, the experimental observations were analyzed in terms of simplified cylindrical electronic systems, and the potential induced by the back-gate voltage, which also removes the cylindrical symmetry, was neglected. Likewise, theoretical calculations dealing with radial electronic systems with an axial magnetic field usually assume a cylindrical symmetry.\cite{TserkovnyakPRB06,BringerPRB11,GladilinPRB13,RosdahlPRB14,RosdahlNL15}
Ferrari {\it et al.}\cite{FerrariNL09} investigated the effect of an axial magnetic field in prismatic systems, but the single-particle model adopted did not allow for a direct comparison with experiments. 

Such approximations are particularly severe in radial heterostructures, where coupling between the discrete (hexagonal in InAs or GaAs) symmetry and many-electron interactions leads to strongly inhomogeneously distributed electron gas and, in turn, to the coexistence of 1D and 2D channels at the corners and facets of the hexagonal hetero-interfaces.\cite{WongNL11,BertoniPRB11,FunkNL13} Strong anisotropy-induced effects are predicted in this case, such as negative magneto-resistance in a transverse magnetic field\cite{RoyoPRB13} and symmetry-induced cancellation of the AB effect in hexagonal quantum rings\cite{BallesterEPL13} The inhomogeneous electron gas localization was crucially exposed in the recent observation of intra- and inter-band excitations.\cite{FunkNL13,JadczakNL14} 

In this paper we study the electronic states and magnetoconductance in GaAs/InAs core-shell NWs with an axial magnetic field within a spin-density-functional theory (SDFT) approach. Our fully 3D modeling explicitly includes the description of the quantum states within an envelope function approach with realistic cross-section and composition of the sample, and includes the electrostatic field induced by external gates in two different device geometries, gate-all-around and back-gate. 
At low magnetic fields, we investigate the nature of the magnetoconductance oscillations, as measured in Ref.~\onlinecite{GulPRB14}, predicting specific signatures of the discrete symmetry of the electronic system in the AB magnetoconductance oscillations, and justifying the observation of AB oscillations despite the broken symmetry induced by the back-gate voltage. 
At high magnetic fields we found that several charge and spin transitions occur. We discuss the origin of these transitions in terms of different magnetic field induced localization and Coulomb regimes and predict their signatures in magnetoconductance experiments.

\section{Theoretical model \label{theory}}

Within a parabolic single-band envelope-function description, the effective Kohn-Sham Hamiltonian under an external magnetic field reads

\begin{equation}
\hat{H}=\frac{1}{2}\left(\hat{\mathbf{P}} - e\, \mathbf{A}(\mathbf{R})\right) \frac{1}{m^*(\mathbf{R})}
\left(\hat{\mathbf{P}} - e \,\mathbf{A}(\mathbf{R})\right) + V_{conf}(\mathbf{R}) + V_{Z}^{\sigma}(\mathbf{R}) + V_H(\mathbf{R}) + V_{XC}^{\sigma}(\mathbf{R}).
\label{eq1}
\end{equation}

\noindent Here, $\mathbf{R}=(x,y,z)$, $\hat{\mathbf{P}}$ is the momentum operator, $\mathbf{A}(\mathbf{R})$ is the vector potential, $e$ is the elementary charge, and $m^*(\mathbf{R})$ is the material-dependent electron effective mass. $V_{conf}(\mathbf{R})$ is the spatial confinement potential induced by the heterostructure and $V_H(\mathbf{R})$ is the Hartree potential energy. The Zeeman energy $V_{Z}^{\sigma}(\mathbf{R})$  and the exchange-correlation potential $V_{XC}^{\sigma}(\mathbf{R})$ depend on the the spin index $\sigma = \uparrow,\downarrow$ of the electrons.

We consider an infinitely long NW extending along the $z$ direction. To describe an axial magnetic field we adopt the symmetric gauge $\mathbf{A}(\mathbf{R})=B/2(-y,x,0)$ (see Fig.~1 for axis definition). The axial field does not break the spatial invariance along the $z$-axis. 
Therefore, the single-particle eigenfunctions of (\ref{eq1}) can be written as $\Psi_{n,k,\sigma}(\mathbf{R})= e^{ikz} \phi_{n,\sigma}(\mathbf{r})$, with $\mathbf{r}\equiv(x,y)$,  $n$ the principal quantum number and $k$ the wavenumber along direction $z$. Substituting $\Psi_{n,k,\sigma}(\mathbf{R})$ and $\mathbf{A}(\mathbf{R})$ in (\ref{eq1}) we obtain the spin-dependent Kohn-Sham equation

%\begin{equation}
\begin{align}
 \bigg[ -\frac{\hbar^2}{2}\nabla_{\mathbf{r}}\frac{1}{m^*(\mathbf{r})}\nabla_{\mathbf{r}} & + \frac{e\,B}{2\,m^*(\mathbf{r})}
\hat{L}_z + \frac{e^2\,B^2}{8\,m^*(\mathbf{r})}\left(x^2 + y^2\right) + v_{conf}(\mathbf{r})  \nonumber\\
+ v_Z^{\sigma}(\mathbf{r}) & + v_H(\mathbf{r}) 
+ v_{xc}^{\sigma}(\mathbf{r})\bigg] \phi_{n,\sigma}(\mathbf{r})= \epsilon_{n,k,\sigma} \, \phi_{n,\sigma}(\mathbf{r}).
 \label{eq2}
\end{align}
%\end{equation}

\noindent Here, $\epsilon_{n,k,\sigma}= \varepsilon_{n,\sigma} + \frac{\hbar^2\, k^2}{2 m^*_z}$ includes the 1D parabolic 
dispersion along the z-axis, and $\hat{L}_z=-i\,\hbar \left(x\,\frac{\partial}{\partial y} - y\,\frac{\partial}{\partial x}\right)$ is the azimuthal angular momentum operator. To obtain Eq.~(\ref{eq2}) it is necessary to assume that the $z$-component of the effective mass, $m^*_z$, does not depend on $\mathbf{r}$, i.e., on the material. This approximation is expected to have a small effect\cite{AndoJPSJ82} and enables to decouple the electron motion in the longitudinal and transverse directions.

The confinement potential $v_{conf}(\mathbf{r})$ is set by the conduction band offsets among the different materials that are radially modulated in the NW cross section. The  Zeeman term is 

\begin{equation}
v_Z^{\sigma}(\mathbf{r})=g^*(\mathbf{r})\,\mu_{B}\,B\,\eta_\sigma\:,
\label{eq3}
\end{equation}

\noindent where $g^*(\mathbf{r})$ is the material dependent Land\'e factor, $\mu_{B}$ is the Bohr magneton, and $\eta_\sigma=+1/2 (-1/2)$ for $\sigma=\uparrow (\downarrow)$. 

The Hartree potential energy, $v_H(\mathbf{r})$, is calculated from the electrostatic potential, $v_H(\mathbf{r})=-e\,\Phi(\mathbf{r})$, via the Poisson equation 

\begin{equation}
\nabla\varepsilon(\mathbf{r})\nabla \Phi(\mathbf{r})=\frac{1}{\varepsilon_0}e(n(\mathbf{r})-n_D(\mathbf{r}))\:.
\label{eq4}
\end{equation}

\noindent Here, $n(\mathbf{r})=n_{\uparrow}(\mathbf{r})+n_{\downarrow} (\mathbf{r})$ is the total free electron charge density calculated, using the Khon-Sham eigenstates obtained from Eq.~(\ref{eq2}), as

\begin{equation}
n_{\sigma}(\mathbf{r})= \frac{1}{2\,\pi} \sum_n \left|\phi_{n,\sigma}(\mathbf{r})\right|^2 \int_{-\infty}^{\infty} dk\, f(\epsilon_{n,k,\sigma}-E_F,T),
\label{eq6}
\end{equation}

\noindent where 

\begin{equation}
f(\epsilon_{n,k,\sigma}-E_F,T)=\frac{1}{1+e^{(\epsilon_{n,k,\sigma}-E_F)/k_B\,T}},
\label{eq7}
\end{equation}

\noindent is the Fermi occupation, with $E_F$, $T$ and $k_B$ being, respectively, the Fermi energy, temperature, and Boltzmann constant. In Eq.~(\ref{eq4}) $n_D(\mathbf{r})$ is the density of static donors and $\varepsilon(\mathbf{r})$ is the material dependent static dielectric constant. 

The exchange and 
correlation potential, $v_{xc}^{\sigma}(\mathbf{r})$, in the local-spin-density approximation (LSDA) is given by 
the functional derivative

\begin{equation}
 v_{xc}^{\sigma}(\mathbf{r})= \frac{\delta \, \varepsilon_{xc}\left(n(\mathbf{r}),\zeta(\mathbf{r})\right)}
 {\delta\,n_{\sigma}(\mathbf{r})}, 
\label{eq5}
\end{equation}

\noindent where $\varepsilon_{xc}\left(n(\mathbf{r}),\zeta(\mathbf{r})\right)$ is the exchange and correlation energy density and 
\begin{equation}\label{eq:zeta}
\zeta(\mathbf{r})=\frac{n_{\uparrow}(\mathbf{r})-n_{\downarrow}(\mathbf{r})}{n(\mathbf{r})}
\end{equation}
is the local spin polarization. In the present paper, we use the correlation functional proposed by Perdew and
Wang.\cite{PerdewPRB92}

From the solutions of the Kohn-Sham equations we also obtain the total free energy per unit length from\cite{HirosePRB01}

%\begin{equation}
\begin{align}
E & =\frac{1}{2\,\pi}\sum_{n,\sigma}  \int_{-\infty}^{\infty} dk \, \epsilon_{n,k,\sigma} \, f_{n,k,\sigma}
-\frac{1}{2} e \int d\mathbf{r} \, v_H(\mathbf{r}) \, n(\mathbf{r})- \sum_{\sigma} \int d\mathbf{r} \, 
v_{xc}^{\sigma}(\mathbf{r}) \, n_{\sigma}(\mathbf{r}) \nonumber \\ 
& + \int d\mathbf{r}\, \varepsilon_{xc}\left(n(\mathbf{r}),\zeta(\mathbf{r})\right) 
+ \frac{k_B \, T}{2 \, \pi} \sum_{n,\sigma} 
\int_{-\infty}^{\infty} dk \,\big[ f_{n,k,\sigma} \ln f_{n,k,\sigma} + (1-f_{n,k,\sigma}) \ln (1-f_{n,k,\sigma})\big].
\label{eq8}
\end{align}
%\end{equation}

\noindent Here, the second term on the right hand side is the Hartree energy per unit length with the sign inverted, the fourth term is the exchange and correlation energy per unit length, and the last term is an entropy functional, where $f_{n,k,\sigma} = f(\epsilon_{n,k,\sigma}-E_F,T)$.

Equations (\ref{eq2})-(\ref{eq8}) are solved iteratively until self-consistency is reached, which we consider to occur when two convergence criteria are simultaneously fulfilled in two consecutive iterations: first, the relative variation of the charge density  
is lower than $10^{-4}$ at any point of the discretization domain, and second, the relative variation in total free energy 
per unit length (Eq.~\ref{eq8}) is lower than $10^{-8}$.

\begin{figure}[h!]
\includegraphics[width=0.5\textwidth]{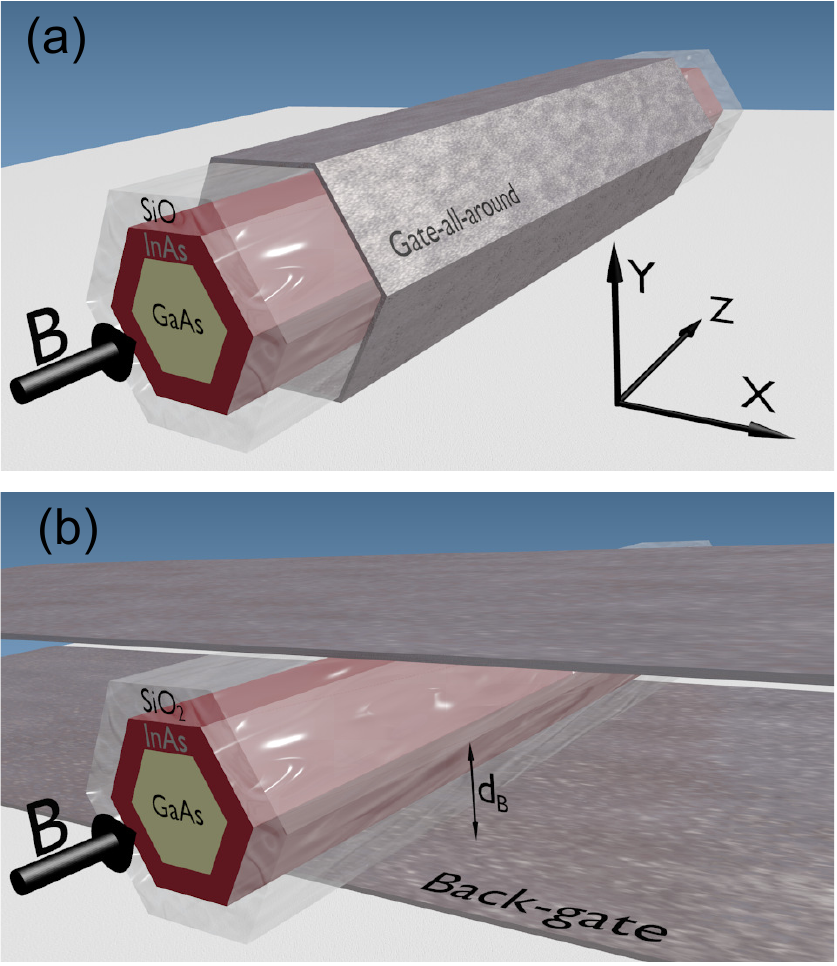}
\caption{Schematics of a core-shell NW in the (a) gate-all-around and (b) back-gate device configurations.}
\label{fig1}
\end{figure}

Equations (\ref{eq2}) and (\ref{eq4}) are numerically integrated in a real space hexagonal domain. 
We use the same symmetry preserving triangular grid with $\sim1.14$ $\mathrm{points/nm^2}$ for both formulas and 
integrate Eqs.~\ref{eq2} and~\ref{eq4} with the methods of finite elements and finite volumes, respectively. 
Dirichlet boundary conditions are assumed in both cases, generally forcing the solutions to vanish at the boundaries. 
To simulate the effect of a gate-all-around (see Fig.~\ref{fig1}(a)), the electrostatic potential in the Poisson equation 
is forced to take the gate voltage $V_g$ at the domain boundaries. For a back-gate, we assume that the hexagonal domain is sandwiched by two flat infinite electrodes (see Fig.~\ref{fig1}(b)) and the electrostatic potential is set at the gate voltage $V_g$ at the bottom facet and zero at the top one. Accordingly, at the lateral boundaries the electrostatic potential is set to 
$F \cdot d_B(\mathbf{r})$, with $F$ and $d_B(\mathbf{r})$ being, respectively, the electric field in the 
capacitor and the vertical distance from the boundary point to the bottom electrode (see Fig.~\ref{fig1}(b)).

Finally, we also calculate the spin-projected free charge density per unit length
\begin{equation}
\bar{n}_\sigma = \int n_\sigma(\mathbf{r}) d\mathbf{r}
\end{equation}
and the spin-projected ballistic conductance of the NW by means of the linear-response Landauer formula
\begin{equation}
G^{\sigma}= \frac{e^2}{h}\sum_n \int_{{\cal B}_{n,\sigma}} -\frac{\partial f(E-E_F,T)}{\partial E} \, dE,
\label{eq9}
\end{equation}
where the integral is performed along each energy spin-subband ${{\cal B}_{n,\sigma}}$. Note that the integrand gives a significant contribution only in the energy region close to the crossings of the subbband with the Fermy energy $E_F$.

\section{Numerical Results \label{Results}}

We consider a GaAs/InAs core-shell hexagonal NW as the one measured in Ref.~\onlinecite{GulPRB14} and outlined in Fig.~\ref{fig1}. It is composed of a GaAs NW core with a minimal diameter of 100 nm and an InAs shell with a thickness of 25 nm. In addition, we include in the device an external 30-nm-thick layer of $\mathrm{SiO_2}$ intended to simulate the insulating layer that separates the conducting channel from a back-gate in the experiment.\cite{GulPRB14} The GaAs core is doped with a homogeneous density of donors $n_D=5\times 10^{15}$ $\mathrm{cm ^{-3}}$. The material parameters used  in the calculations are  listed in Tab.~\ref{table_mat_par}, where the conduction band edge, $E_{CB}$, is obtained with the so-called 40:60 rule~\cite{AdachiBOOK93,DebbarPRB89} from the band gap. Calculations have been conducted assuming a Fermi energy placed 75 meV above the InAs conduction 
band edge (as in Ref.~\onlinecite{GulPRB14}), a temperature of 1.8 K, and the InAs effective mass as the constant mass factor ($m^*_z$) arising in the parabolic dispersion of the 1D subbands (see Eq.~(\ref{eq2})). 

\begin{table}
\begin{tabular}{lccc}
  \hline
  % after \\: \hline or \cline{col1-col2} \cline{col3-col4} ...
          & GaAs  & InAs  & $\mathrm{SiO_2}$ \\
          \hline
          \hline
   $m^*$ & 0.067 & 0.028 & 0.41 \\
   $\varepsilon$ & 13.18 & 15.5 & 3.9 \\
   $g^*$  &  -0.484 & -14.3 & 2.0 \\
   $E_{CB}$ (eV) & 0.858 & 0.252 & 5.4 \\
   \hline
\end{tabular}
\caption{Material parameters used in the simulations; electron effective mass ($m^*$), dielectric
constant ($\varepsilon$), effective Land\'e factor ($g^*$) and conduction band edge ($E_{CB}$). }
\label{table_mat_par}
\end{table}

\subsection{Low magnetic field regime: magnetoconductance oscillations\label{lowfield}}

\begin{figure}[h!]
\includegraphics[width=0.5\textwidth]{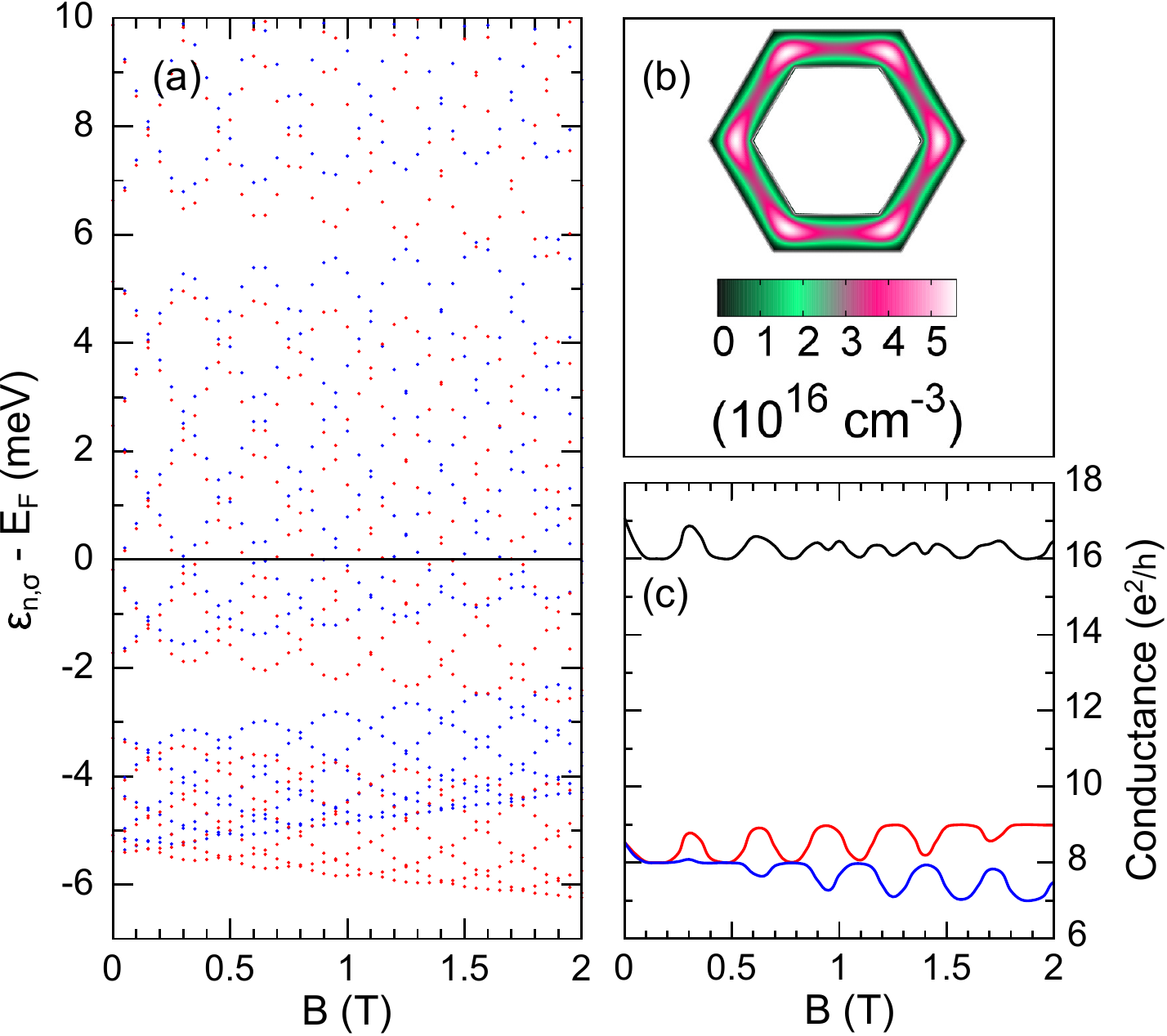}
\caption{(a) Magnetic spin-subbands (MSS) in the low field regime. Red and blue dots indicate \textdownarrow- MSSs and \textuparrow-MSSs, respectively. The horizontal black line is set at $E_F$. (b) Self-consistent electron density distribution, $n(\mathbf{r})$, for the InAs/GaAs NW at $B=0$. (c) Spin-projected magnetoconductances $G_\downarrow$ (red) and $G_\uparrow$ (blue), and total magnetoconductance (black).}
\label{fig2}
\end{figure}

In Fig.~\ref{fig2} we show the ground state properties and magnetoconductance of the investigated core-shell NW at $V_g=0$. The density distribution of conduction band electrons (Fig.~\ref{fig2}(b)) shows that charge is exclusively accumulated in the InAs shell and preferentially localized at the corners of the hexagonal section. As reported for several core-(multi)shell hexagonal NWs,\cite{WongNL11,BertoniPRB11,FunkNL13,MartinezNL12} 
such distribution is favored by Coulomb interactions which tend to increase the inter-electron distance. In Fig.~\ref{fig2}(a) we
show the energies of the spin-subbands edges at different magnetic fields, hereafter referred as magnetic spin-subbands (MSS), with
spin up (\textuparrow-MSSs) and spin down (\textdownarrow-MSSs).
Due to the hexagonal symmetry of the self-consistent potential, the low-energy spectrum is at low fields formed of groups of 12 MSSs arising from the 6 irreducible representations of the $\mathrm{C_{6}}$ symmetry group. Each of these groups is further spin-split by the strong Zeeman effect forming two bunches, one of \textuparrow-MSS and the other of \textdownarrow-MSSs, of 6 braided MSSs. 

Within each group the MSSs oscillate due to the AB effect, developing crossings with 
MSSs of their same group, that have different symmetry and/or different spin, and anticrossings with MSSs of neighbouring groups with same symmetry and spin. The oscillation period is $\sim0.32$ T. Since the calculated expectation value of the radial position, $\rho=\sqrt{x^2+y^2}$, of the electron system is $66.36$ nm, this periodicity fairly corresponds to the periodicity of $\sim 0.30$ T of the corresponding circular system.

In Fig.~\ref{fig2}(c) we show the spin-projected magnetoconductances $G_\sigma(B)$ and the total magnetoconductance  $G(B)=G_\uparrow(B)+G_\downarrow(B)$. Even though both $G_\sigma(B)$ exhibit regular flux-periodic oscillations, $G(B)$ only does so at very low fields. After the second oscillation cycle the $G(B)$ periodicity is suppressed by the Zeeman effect which breaks the periodicity of the MSS spectrm.~\cite{TserkovnyakPRB06,RosdahlPRB14} Apart from this, $G(B)$ does not differ qualitatively from that of an electron system in a cylindrical tube.\cite{TserkovnyakPRB06, GulPRB14, RosdahlPRB14} Indeed, in the present case, $E_F$ lies within one group of braided MSSs, and the spectrum around $E_F$ is similar to that of a cylindrical system. However, in an experiment $E_F$ can be tuned by means of external gates. Therefore, we next study the system at different Fermi levels $E_F$ or applied gate voltages $V_g$.

\begin{figure}[h!]
\includegraphics[width=0.5\textwidth]{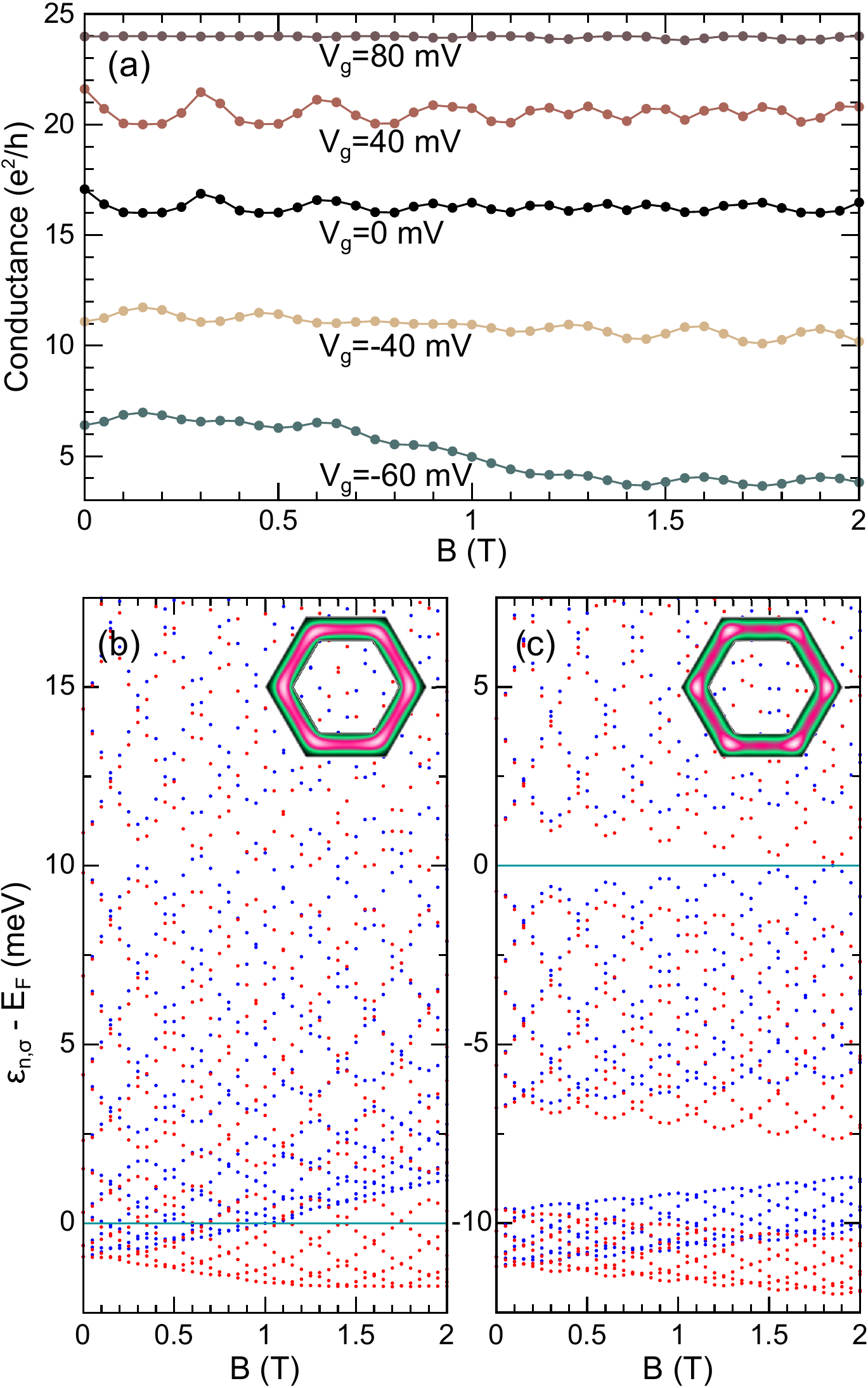}
\caption{(a) Total magnetoconductance  at selected gate-all-around voltages, $V_g$, as indicated by labels (the $V_g=0$ curve is the same as the black line in Fig.~\ref{fig2}(c)). (b) MSSs at $V_g = -60 \mbox{mV}$. (c) MSSs at $V_g = 80 \mbox{mV}$. Insets in (b), (c) show the corresponding $n(\mathbf{r})$.}
\label{fig3}
\end{figure}

In Fig.~\ref{fig3} we show the effect of a gate-all-around voltage. This geometry tunes the position of the MSSs with respect to $E_F$, modulating the total density in the system while preserving the hexagonal symmetry. As shown in Fig.~\ref{fig3}(a), the oscillatory behavior of the magnetoconductance due to the AB effect is absent at certain voltages. For instance, at $\mathrm{V_g=80}$ mV the magnetoconductance is flat. This is due to the positioning of $E_F$ in the energy gap between the second and third group of MSSs, as shown in Fig.~\ref{fig3}(c). Since $E_F$ does not cross any MSS, the number of conducting channels is constant. Comparing Figs.~\ref{fig3}(b),(c), which correspond to $\mathrm{V_g}$=-60 and 80 mV, respectively, we also observe that the gate voltage affects both the width of the MSSs groups and the gaps between them. In fact, $V_g$ affects the total electron density and, hence, electron localization. As shown in the insets of Figs.~\ref{fig3}(b),(c), a $V_g>0$ favors localization in the corners of the InAs shell, due to the larger electron-electron interaction. This, in turn, reduces the tunneling among states at the corners and, hence, the splittings within bunches of MSSs, while it increases the gaps between consecutive bunches.\cite{FerrariNL09} Note that, since the latter gaps are a direct consequence of the discrete symmetry of the system, flat magnetoconductance  is a direct signature of the hexagonal symmetry, which is more likely to be observed at positive gate voltages.

Observation of flat magnetoconductance when sweeping $V_g$ has not been reported in the transport measurements performed hitherto on hexagonal NWs under axial magnetic fields.\cite{GulPRB14,GulNL14,WenzAPL14,BlomersNL11} However, in these works the electron density was normally modulated by a back-gate instead of a gate-all-around. The electrostatic field generated by a back-gate removes the hexagonal symmetry of the electronic system and it could even destroy the doubly-connected topology that originates the AB effect. Therefore, one may wonder why flux-periodic oscillations in the magnetoconductance are observed at all.

\begin{figure}[h!]
\includegraphics[width=0.5\textwidth]{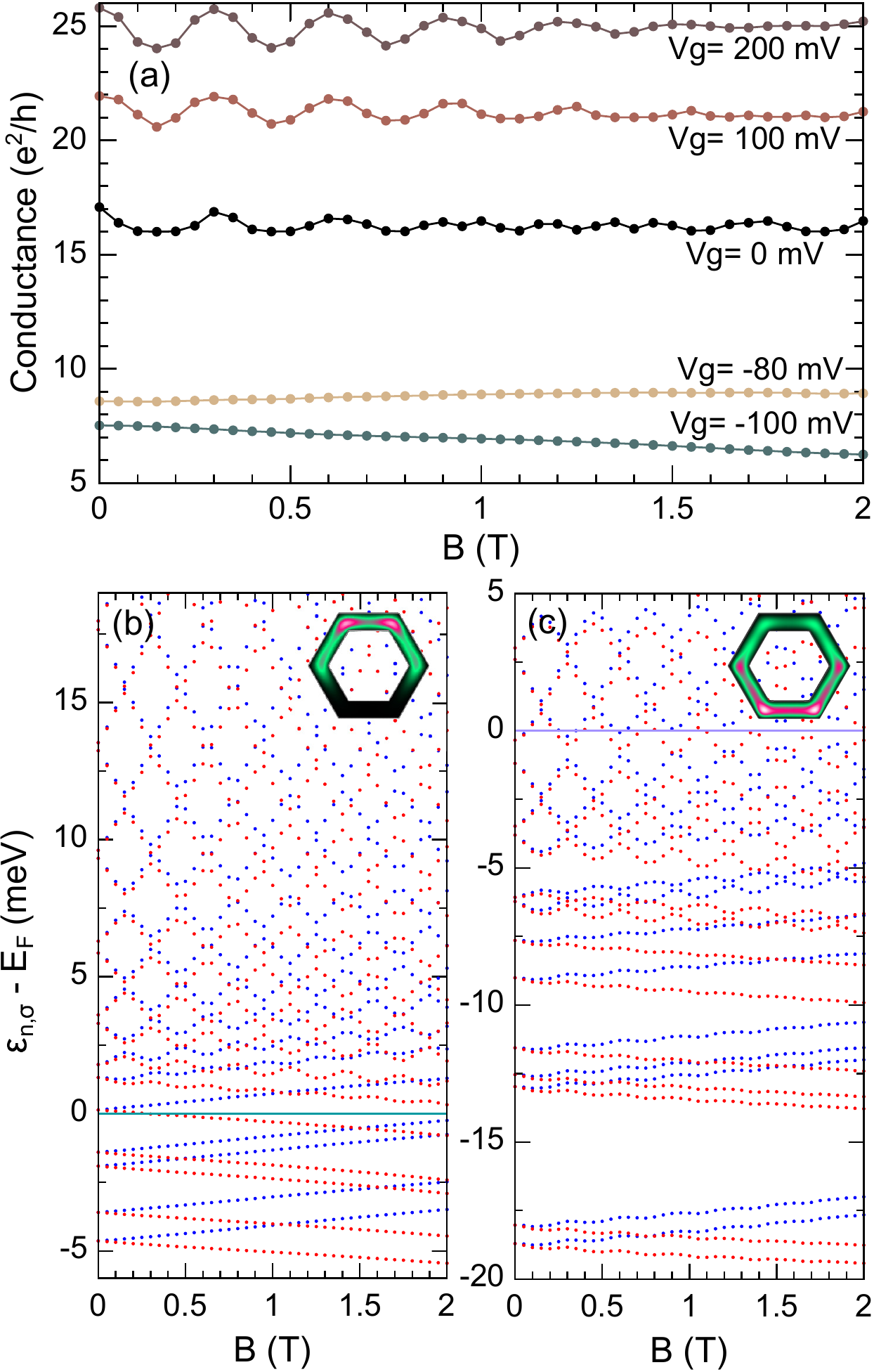}
\caption{(a) Same as in Fig.~\ref{fig3} but for a back-gate device. (b) MSSs at $V_g = -80 \mbox{mV}$ (c) MSSs at $V_g = 200 \mbox{mV}$.}
\label{fig4}
\end{figure}

In order to asses this point, in Fig.~\ref{fig4} we show the results of simulations performed at 
different back-gate voltages. As shown in the insets of Figs.~\ref{fig4}(b),(c), the applied voltage strongly reshapes the electron density distribution in the NW. At negative (positive) $V_g$ the total density in the system is reduced (increased) and concentrated in the top (bottom) half of the InAs shell. However, whereas the doubly-connected topology which results in AB oscillations is removed at $V_g<0$, it is robust for $V_g>0$. The origin of this difference can be appreciated from the corresponding MSSs (Figs.~\ref{fig4}(b),(c)). The lowest lying MSSs are strongly affected by the gate, loosing the doubly-connected topology and showing an almost linear dispersion with the magnetic field. Higher energy MSSs, on the contrary, being more delocalized over the NW section, still show doubly-connected topology. Since at $V_g<0$ only low-lying MSSs are occupied (see Fig.~\ref{fig4}(b)), the total electron density loses the doubly-connected topology and the corresponding magnetoconductance does not show AB oscillations. By contrast, at $V_g>0$ several states with the doubly-connected topology are occupied and the AB oscillations of the magnetoconductance persist (see Fig.~\ref{fig4}(a)). The latter is indeed the usual regime in magnetotransport experiments\cite{GulPRB14, GulNL14} where, therefore, periodic oscillations in the magnetoconductance are observed despite the symmetry reduction.

\subsection{High magnetic field regime: spin and charge transitions \label{highfield}}

\begin{figure}[h!]
\includegraphics[width=0.5\textwidth]{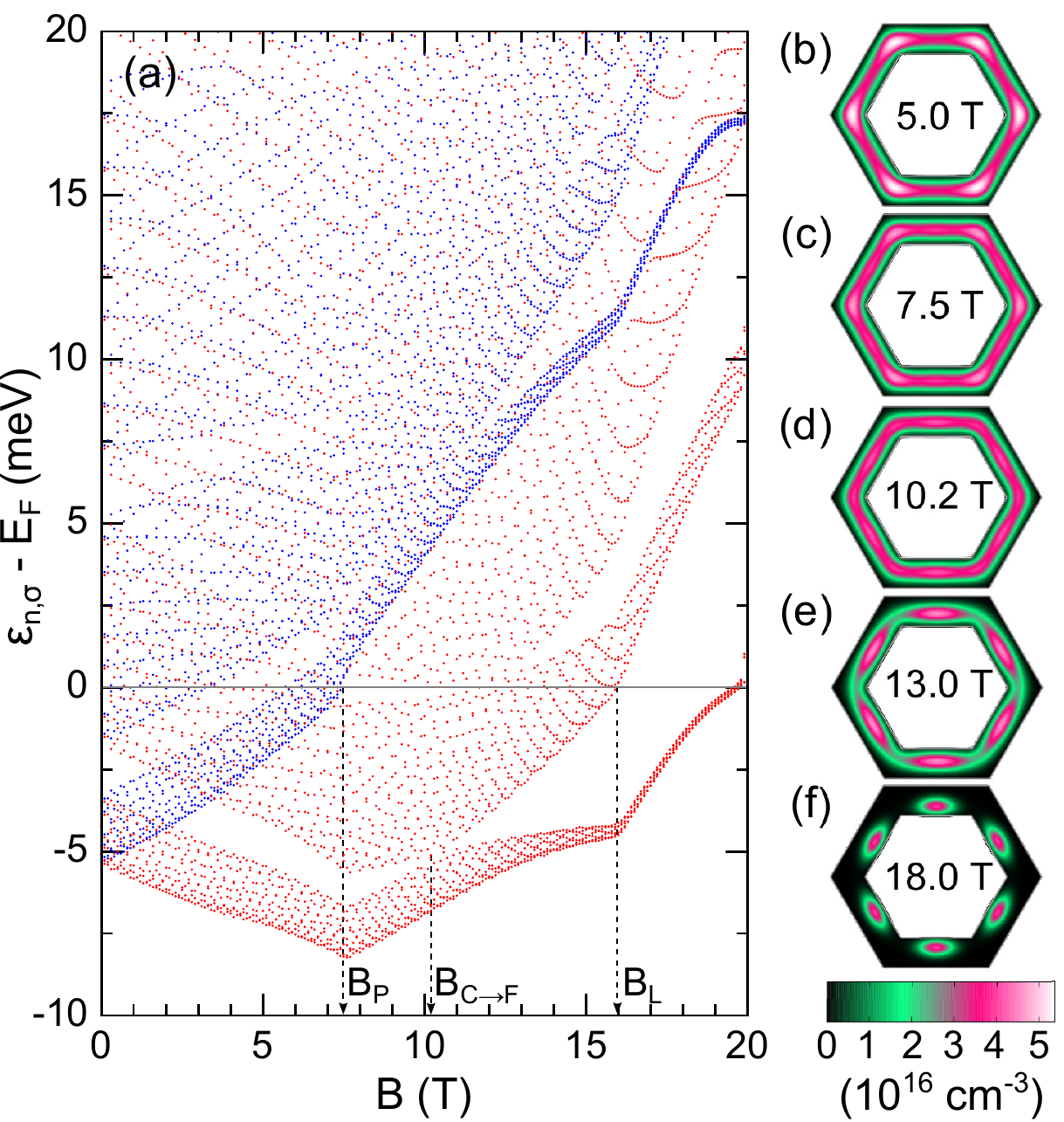}
\caption{(a) MSSs up to complete charge depletion. Blue (red) dots are used for \textuparrow-MSSs (\textdownarrow-MSSs). The horizontal line indicates the position of $E_F$. Vertical dashed arrows indicate fields at which different spin/charge transitions occur (see text). (b)-(f) Self-consistent electron density distributions $n(\mathbf{r})$ at selected magnetic fields.}
\label{fig5}
\end{figure}

We next study the high magnetic field regime, up to the limit of complete electron depletion, which occurs at $B\sim 20$ T in this sample. Fig.~\ref{fig5} shows the MSSs and the self-consistent total electron density distributions at selected fields (spin-projected electron densities show only minor differencies and are not shown here). All simulations in this section are performed at $V_g=0$. The overall behaviour of MSSs shows that, in addition to the diamagnetic shift, several transitions occur at discrete fields, as we discuss below.

The evolution of $n(\mathbf{r})$ in Figs.~\ref{fig5}(b)-(f) shows that the axial field induces a transition from an electron distribution localized at the corners (low field, Figs.~\ref{fig5}(a),(b)) to a distribution increasingly localized in the center of the facets (high field, Figs.~\ref{fig5}(e),(f)). This charge reshaping is induced by the diamagnetic term (third term in the left hand side of Eq.~\ref{eq2}), which constrains the electron density to adopt distributions with lower radius as the field is increased, counteracted by Coulomb interactions.

Such a corner-to-facet transition can be correlated with the evolution of the MSSs. In Fig.~\ref{fig5}(a) the lowest lying bunch of 12 MSSs at $B=0$ corresponds to states localized at the corners, whereas the second set of states are localized at the facets for orthogonality. As the field is increased Zeeman spin-splitting takes place and the two sets of 6 \textdownarrow-MSSs approach in energy eventually overlapping at $\mathrm{B_{C\rightarrow F}} ~ \mathrm{\sim 10.2\, T}$. At this point, the 2D electron density integrated along the minimal (facet-to-facet) and maximal (corner-to-corner) diameter\cite{BertoniPRB11} is nearly the same (see Fig.~\ref{fig5}(d)). At $B>\mathrm{B_{C\rightarrow F}}$ the 6 lowest \textdownarrow-MSSs are localized at the facets of the inner interface whilst corner states are much higher in energy, corresponding to the third group of 6 \textdownarrow-MSS. The same transition occurs for \textuparrow-MSS, however, these states are already depopulated at the transition field.

%A comparison with a non-interacting simulation highlights the many-electron origin of this transition.  The MSSs with $v_H=0, v_{XC}=0$ are shown in Fig.~\ref{fig9}(a). The position of $E_F$ has been chosen to yield the same total electron density as the full calculation of Fig.~\ref{fig5} at $B=0$. Clearly MMSs show a smooth evolution and no signature of transition is present. Furthermore, the self-consistent charge density at increasing field (Figs.~\ref{fig9}(b)-(f) show that the charge is delocalized over the ring at zero field and localization does not change qualitatively with increasing fields.

\begin{figure}[h!]
\includegraphics[width=0.5\textwidth]{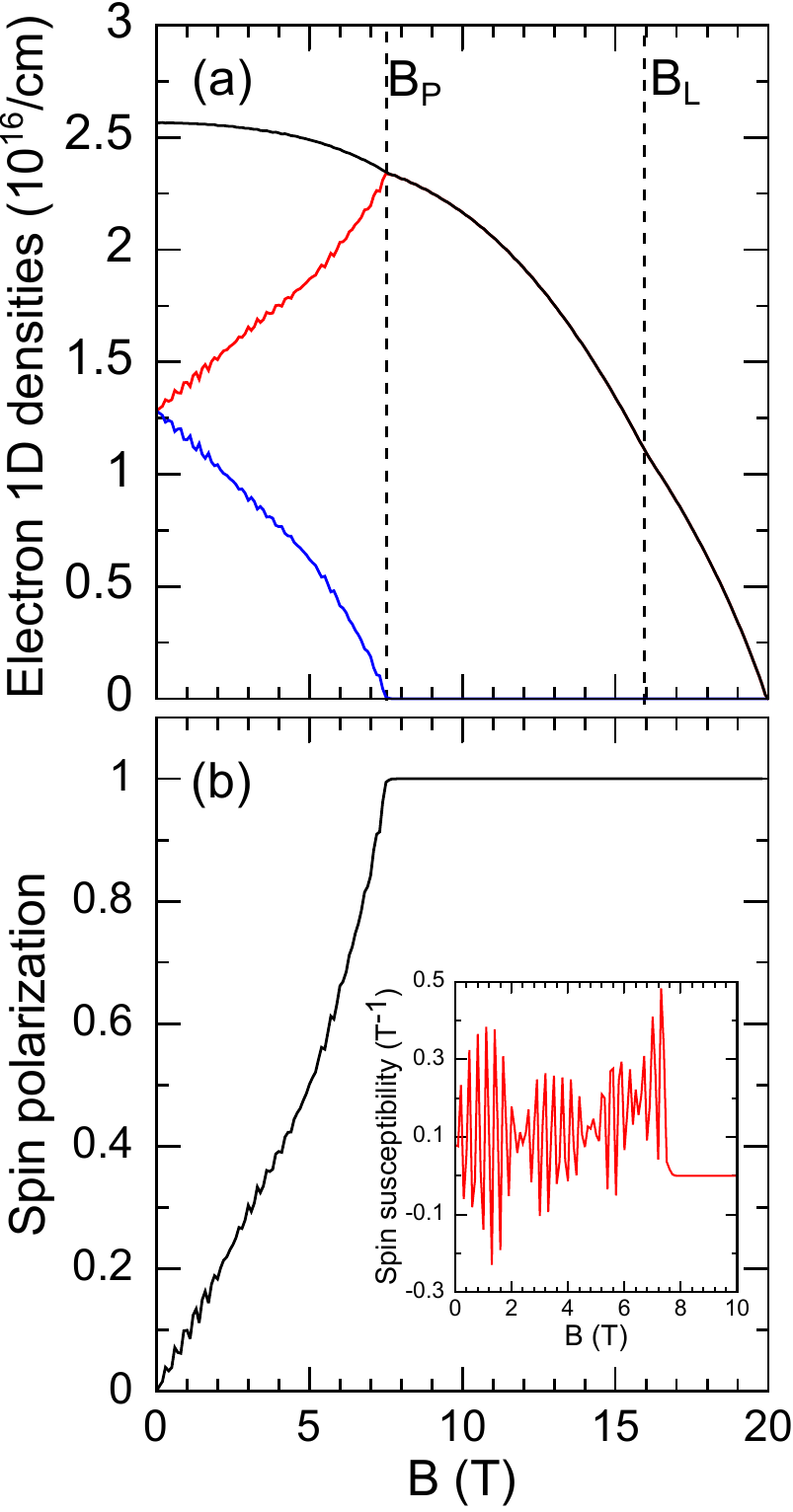}
\caption{(a) Total density $\hat{n}$ (black lines) and spin-projected densities $\bar{n}_\uparrow$ (blue lines) and $ \bar{n}_\downarrow$ (red lines) as a function of the field intensity $B$. Vertical dashed lines illustrate the transition fields in Fig.~\ref{fig5}(a). (b) Spin polarization as a function of the magnetic field. Inset: spin susceptibility.}
\label{fig6}
\end{figure}

Apart from this smooth spatial localization transition, two abrupt changes of slope appear in the calculated MSSs.
The first one occurs at $\mathrm{B_P}=7.5\: \mbox{T}$ and corresponds to complete spin polarization, as demonstrated by the spin-projected electron densities $\bar{n}_\uparrow, \bar{n}_\downarrow$ shown in Fig.~\ref{fig6}(a) and the corresponding spin polarization in Fig.~\ref{fig6}(b), which marks a clear transition to a ferromagnetic state at $B_P$. Note that the total density (black line in Fig.~\ref{fig6}(a)) is reduced by the magnetic field with a roughly parabolic trend due to the depletion of successive, high energy MSSs. However, the curve shows a change of slope at ($\mathrm{B_P}$). At fields right after $B_P$ the rate at which the NW is depleted decreases momentarily. $\bar{n}_\downarrow$ passes abruptly from being increased to decreased at $B_P$, in agreement with the inversion of the \textdownarrow-MMSs slope exposed in Fig.~\ref{fig5}(a).

The singular behavior of the spin polarization (Fig.~\ref{fig6}(b)) is reminiscent of the first-order phase transition of a 2D electron gas with an in-plane magnetic field\cite{SubasiPRB08,ZhangPRL06} (note that in our system the Seitz radius $r_s\sim 0.07$ at zero field, which is a very weakly correlated regime), although it is difficult in our numerical treatment to establish whether it is a weakly first order or continuous transition. The inset in Fig.~\ref{fig6}(b) shows the spin susceptibility, i.e., the magnetic field derivative of the spin polarization. This magnitude oscillates with the field as a consequence of the interplay between the AB effect and the Zeeman splitting which produce short period modulations of the spin densities. 

%Note that the total density (black line in Fig.~\ref{fig6}(a))is reduced by the magnetic field with a roughly parabolic trend due to the depletion of successive, high energy MSSs. However, the curve shows a change of slope at ($\mathrm{B_P}$). For $B>B_P$ the rate at which the NW is depleted, decreses. $\hat{n}_\downarrow$ increases for $B<B_P$, therefore the polarization takes place at a larger rate than depletion. This is the origin of the large change of slope occurring for \textdownarrow-MSSs in Fig.~\ref{fig5} at this field. Indeed, for $B<B_P$ the gain in Zeeman energy overcomes the loss in Hartree energy due to depletion, while for $B>B_P$ the opposite holds.

%The single-particle description completely changes this picture. The corresponding charge densities and polarizations are also shon in Fig.~\ref{fig7}. The charge density is depleted by the field at a much faster rate than in SDFT. As a result, the population of \textdownarrow electrons starts decreasing before polarization takes place and the polarization transition takes place at much smaller field.\footnote{In Ref.~\onlinecite{SubasiPRB08} the transition for interacting electrons takes place at lower field than for the non-interacting system, contrary to our case. However in those calculations $\hat{n}$ is kept fixed, while in our case it is self-consistently determined and decreasing with increasing field.} 

\begin{figure}[h!]
\includegraphics[width=0.5\textwidth]{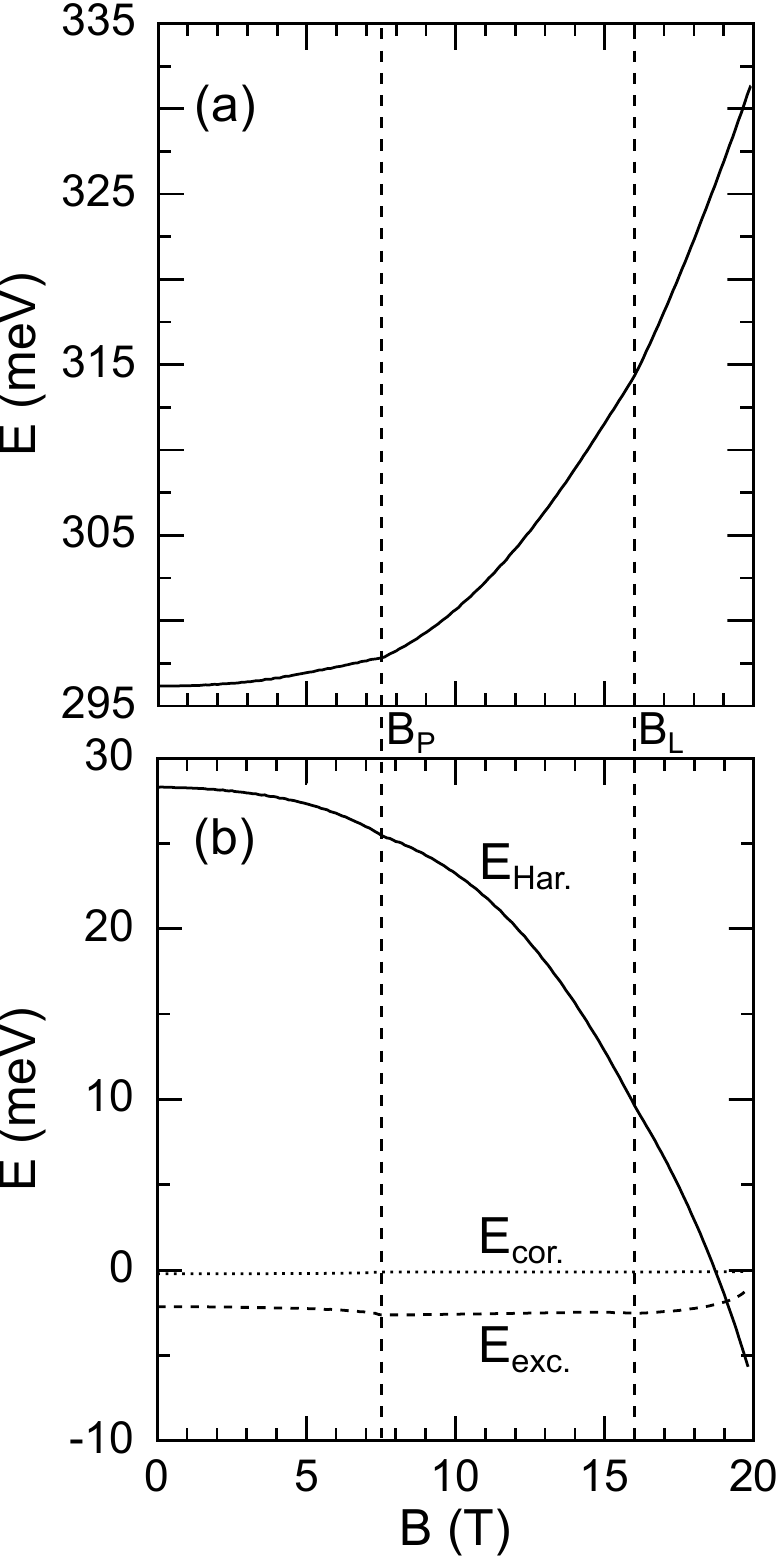}
\caption{(a) Total free energy per electron and (b) Hartree (solid), exchange (dashed) and correlation (dotted) energies per electron as a function of the field $B$. Vertical dashed lines indicate the transition fields in Fig.~\ref{fig5}.}
\label{fig7}
\end{figure}

At fields higher than $B_P$ and $\mathrm{B_{C\rightarrow F}}$ the MSSs shown in Fig.~\ref{fig5}(a) rearrange in groups of six which tend to form Landau-like bands. Finally, at a larger field $B_L=16$ T the spectrum shows an additional transition. This corresponds to complete depletion of the incipient second Landau-like band. The transition is also marked by a weak but visible kink in $\bar{n}(B)$, as shown in Fig.~\ref{fig6}(a), which, as for the ferromagnetic transition, indicates a momentary decrease of the depletion rate. 

The free energy per electron and the many-electron energy contributions per electron are calculated dividing the corresponding magnitudes per unit length by the total electron density and plotted in Figs.~\ref{fig7}(a) and (b), respectively. All energy contributions show weak kinks at $\mathrm{B_P}$ and $\mathrm{B_L}$. The free energy per electron increases with B due to the increase in magnetic confinement. However, the Hartree energy per electron (Fig.~\ref{fig7}(b)) is reduced with B due to field-induced charge depopulation. At high magnetic fields, $\mathrm{B>18.5}$ T, the Hartree energy changes sign because the free electron density is lower than the total density of static donors included in the simulation in the NW GaAs core. Note from Fig.~\ref{fig7}(b) that the direct Hartree energy is one and two orders of magnitude larger than the exchange and correlation contributions, respectively, and therefore, it will rule many-electron effects in the system.

\begin{figure}[h!]
\includegraphics[width=0.5\textwidth]{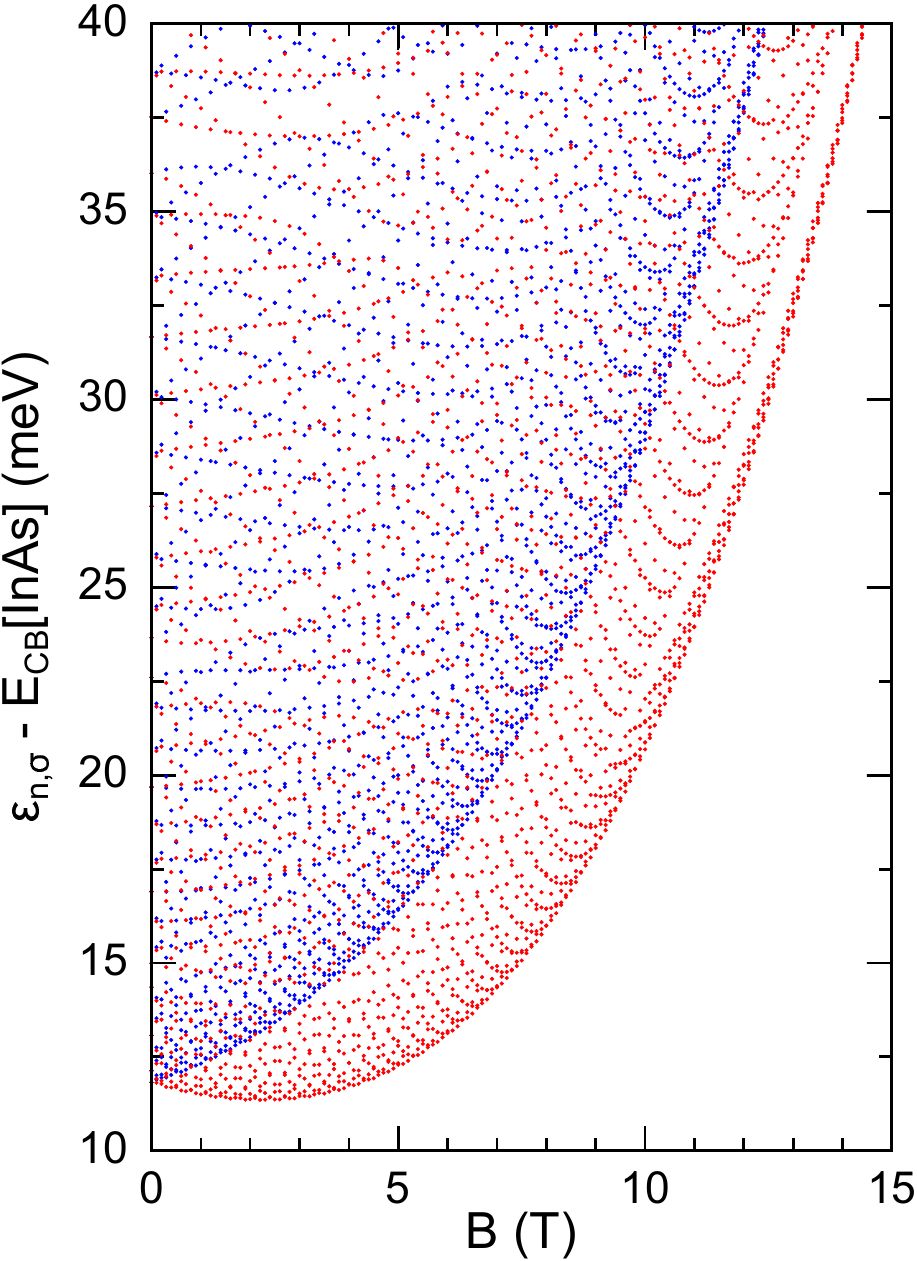}
\caption{Non-interacting MSSs edges with respect to the InAs conduction band edge. Blue (red) dots are used for \textuparrow-MSSs (\textdownarrow-MSSs)}
\label{fig9}
\end{figure}

In order to asses the role of many-electron contributions, in Fig.~\ref{fig9} we show the MSSs calculated in a non-interacting model, i.e., $v_H=0$ and $v_{XC}=0$. The MSSs follow in this case a smooth evolution with B, which evidences the many-electron origin of the two transitions at $B_P$ and $B_L$ in the SDFT calculation. We have also checked that such transitions persist when only $v_{XC}=0$ (data not shown here) as was expected from the weak effect of the exchange and correlation contributions in the present system (see Fig.~\ref{fig7}(b)).

Indeed, the transitions at $\mathrm{B_P}$ and $\mathrm{B_L}$ result from the balance between the two main energy contributions: the magnetic confinement, which increases the system energy with B, and the direct Coulomb or Hartree energy, which is reduced with B due to the charge depletion decreasing in this way the system energy. Thus, the first transition at $\mathrm{B_P}$, which produces an inversion in the slope of the \textdownarrow-MSSs, can be understood as a transition between a regime, $\mathrm{B<B_P}$, in which the reduction in Hartree energy dominates over the magnetic confinement, to another regime,  $\mathrm{B>B_P}$, in which the magnetic confinement dominates. The key difference before and after $\mathrm{B_P}$ is the magnitude
of the Hartree energy that is lost per depleted state, which is larger at $\mathrm{B<B_P}$. This is because when the system is not spin-polarized the Hartree energy also arises from the interactions between electrons with anti-parallel spin. The latter, which are absent in the ferromagnetic phase, are stronger than interactions between parallel spin electrons due to the lack of Fermi hole.

The transition at $\mathrm{B_L}$, which produces an abrupt increase of the MSSs, is also interpreted with similar arguments, i.e., the Hartree energy lost per depleted state is lower at $\mathrm{B>B_L}$. This is due to the larger localization of the electron density at $\mathrm{B>B_L}$ (cf. Figs.~\ref{fig5}(e) and (f)), which entails a larger Fermi hole in the direct Coulomb interaction in this regime. Indeed, it has been proved that the conditional probability of finding an electron with a given spin when there is already another electron with the same spin nearby is lower when the former is localized.\cite{MatitoJCP06}

\begin{figure}[h!]
\includegraphics[width=0.5\textwidth]{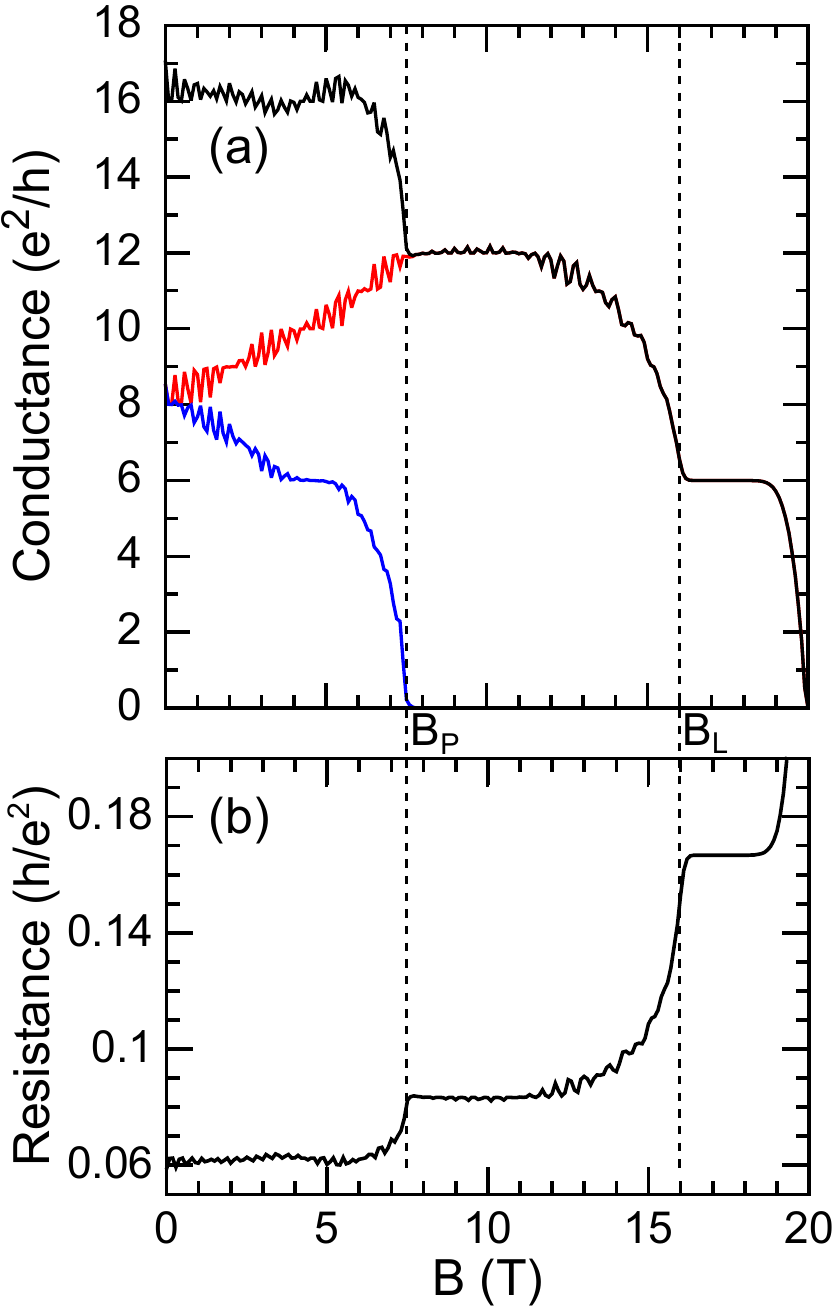}
\caption{(a) Total magnetoconductance  $G=G_\uparrow+G_\downarrow$ (black) and spin resolved magnetoconductances, $G_\uparrow$ (blue) and $G_\downarrow$ (red). (b) Total magnetoresistance. Vertical dashed lines indicate the transition fields in Fig.~\ref{fig5}.}
\label{fig8}
\end{figure}

The spin-projected magnetoconductances $G_\uparrow, G_\downarrow$ and the total magnetoconductance $G=G_\uparrow+G_\downarrow$ calculated from the SDFT modeling are shown in Fig.~\ref{fig8}(a). Starting from low fields, the total magnetoconductance oscillates, due to oscillating MSSs crossing $E_F$, around an average value of 16 $\mathrm{e^2/h}$ up to a magnetic field $\mathrm{B\sim6}$ T. As the field approaches $B_{P}$, a sudden step-like reduction of 4 magnetoconductance units is caused by the sudden depletion of the lowest set of \textuparrow-MSSs (see Fig.~\ref{fig5}(a)) induced by the ferromagnetic transition. 

At $B>B_P$ the magnetoconductance shows an almost flat plateau that lasts up to $\mathrm{B\sim12}$ T. This originates in the location of $E_F$ in the symmetry induced energy gap between the second and third group of six \textdownarrow-MSSs (see Fig.~\ref{fig5} (a)). As $E_F$ merges in the second group of \textdownarrow-MSSs the magnetoconductance starts to oscillate again, while reducing in average at an increasing rate approaching $B_{L}$. At $B>B_{L}$ $E_F$ lies in the wide energy gap between the first and second Landau bands, hence $G$ is constant. 
%It is worth noting that in this regime \textdownarrow-spin electrons are solely transported through six decoupled quasi-1D channels localized at the center of the facets of the InAs/GaAs heterojunction. However, the coupling/tunneling among these channels can be increased by reducing the magnetic field below $\mathrm{B_L}$ (cf. electron density profiles in Fig.~\ref{fig5} (e) and (f)).  

Finally, $G(B)$ drops to zero when the first incipient Landau band crosses $E_F$ and the conduction band gets completely depleted. In Fig.~\ref{fig8}(b) we also plot the magnetoresistance $1/G(B)$ to illustrate the kink observed at $B_{P}$, which corresponds to that observed in experimental measures\cite{PiotPRB09,TutucPRB03} of ferromagnetic transitions in flat \emph{quasi}-2D electron systems under in-plane magnetic fields. 

\section{Summary and conclusions \label{conc}}

We  performed a SDFT study of the electronic structure and magnetoconductance of hexagonal core-shell NWs pierced by an axial magnetic field. Critically, our modeling goes beyond often employed cylindrical and/or single-particle approximations 
to simulate radial hetero-structures, which neglect the strongly inhomogeneous, field-dependent distribution of the electron gas. 

In the low field regime ($\mbox{B} \lesssim 2 \mbox{T}$) we predict that AB magnetoconductance oscillations may disappear/resurface as a function of the gate-all-around voltage as a direct consequence of the presence of discrete symmetry induced energy gaps. Our calculations also allowed to critically analyze recent experiments~\cite{GulPRB14,GulNL14} and justify the observation of AB oscillations in spite of the broken symmetry induced by the back-gate voltage. 

In the high magnetic field regime we found several field-induced transitions. First, the diamagnetic confinement induces a strong reshaping of the electron gas, which goes through a smooth transition from a low-field electron density distribution concentrated in the corners to a high-field distribution strongly localized in the facets of the radial heterojunction. Several experimental consequences of such reshaping are expected, for example in optical recombination experiments, due to the different localization of electrons and holes.\cite{JadczakNL14}

In addition, two abrupt transitions occur at discrete fields which are related to depletion of higher MSSs, either of the lowest anti-parallel spin MSSs, leading to spin polarization, or of the second incipient Landau-like band with parallel spin. The origin of these transitions lies in the increase of the effective Fermi hole occurring at each transition which affects the amount of Hartree energy that is lost per depleted state. As a consequence, such abrupt transitions are clearly marked in the calculated magnetoconductance by step-like behaviors.

\begin{acknowledgments}
We acknowledge partial financial support from Universitat Jaume I - Projects No. P1-1B2011-01 and P1.1B2014-24, MINECO Project No. CTQ2011-27324, APOSTD/2013/052 Generalitat Valenciana Vali+d Grant (MR), and MINECO FPU Grant (CS), from EU-FP7 Initial Training Network INDEX, and from University of Modena and Reggio emilia, through Grant “Nano- and emerging materials and systems for sustainable technologies”.
\end{acknowledgments}

\bibliography{NW_axial_field}

\end{document}